\documentclass[prl,aps,twocolumn,showpacs,amsmath,amssymb]{revtex4}
\usepackage{epsfig}

\newcommand{\CaFeAs}{CaFe$_2$As$_2$}

\newcommand{\AFeAs}{AFe$_2$As$_2$}
\newcommand{\BaNiAs}{BaNi$_2$As$_2$}
\newcommand{\BaNiP}{BaNi$_2$P$_2$}
\newcommand{\ThCrSi}{ThCr$_2$Si$_2$}

\begin{document}

\title{First Order Phase Transition and Superconductivity in BaNi$_2$As$_2$ Single Crystals}

\author{F. Ronning$^1$, N. Kurita$^1$, E.D. Bauer$^1$, B.L. Scott$^1$, T. Park$^{1,2}$,
T. Klimczuk$^{1,3}$, R. Movshovich$^1$, J.D. Thompson$^1$}
\affiliation{$^1$Los Alamos National Laboratory, Los Alamos, New Mexico 87545, USA\\
             $^2$Department of Physics, Sungkyunkwan University, Suwon 440-746, Korea\\
             $^3$Faculty of Applied Physics and Mathematics, Gdansk University of Technology, Narutowicza 11/12, 80-952 Gdansk, Poland}

\date{\today}

\begin{abstract}
We report the synthesis and physical properties of single crystals
of stoichiometric \BaNiAs{} that crystalizes in the ThCr$_2$Si$_2$
structure with lattice parameters $a$ =  4.112(4) \AA{} and $c$ =
11.54(2) \AA{}. Resistivity and heat capacity show a first order
phase transition at $T_0$ = 130 K with a thermal hysteresis of 7 K.
The Hall coefficient is weakly temperature dependent from room
temperature to 2 K where it has a value of -4x10$^{-10}$
$\Omega$-cm/Oe. Resistivity, ac-susceptibility, and heat capacity
find evidence for bulk superconductivity at $T_c$ = 0.7 K. The
Sommerfeld coefficient at $T_c$ is 11.6 $\pm$ 0.9 mJ/molK$^2$. The
upper critical field is anisotropic with initial slopes of
d$H_{c2}^{c}$/d$T$ = -0.19 T/K and d$H_{c2}^{ab}$/d$T$ = -0.40 T/K,
as determined by resistivity.
\end{abstract}


\maketitle

The large superconducting transition temperatures found in the
oxypnictide system has stimulated a great deal of research activity
world-wide. Perhaps, even more remarkable than the large transition
temperatures (up to 55 K for SmFeAs(O,F)\cite{ZARen2008a}) is the
large tunability these systems possess. Superconductivity has been
found in RTPn(O,F) (ZrCuSiAs structure type) at ambient pressure
with Rare-earth R = La, Ce, Pr, Nd, Sm, or Gd; Transition metal T =
Fe or Ni, and Pnictide Pn = P or As
\cite{ZARen2008a,KamiharaJACS2008,XHChenNature2008,GFChen2008a,
ZARen2008b,ZARen2008c,PCheng2008,Watanabe2007LaNiPO,Watanabe2008LaNiAsO,
Fang2008NiAs,Li2008NiAs,Kamihara2006LaFePO}. Some compounds require
chemical substitution, while for others the parent compounds also
superconduct. Furthermore, superconductivity has also been
discovered in the related \ThCrSi{} structure type, where it has
been found by doping \AFeAs{} on the A site (with A = Ba, Sr, Ca,
Eu)
\cite{Rotter2008b,GFChen2008b,Sasmal2008,Wu2008Ca,Jeevan2008EuFe2As2SC},
under pressure in
\AFeAs\cite{Park2008CaFe2As2,Torikachvili2008CaFe2As2,Alireza2008BaFe2As2},
and at ambient pressure in the stoichiometric compounds
\BaNiP\cite{Mine2008BaNi2P2},
LaRu$_2$P$_2$\cite{Jeitschko1987LaRu2P2}, CsFe$_2$As$_2$, and
KFe$_2$As$_2$\cite{Sasmal2008}.

The common structural element between the ZrCuSiAs and \ThCrSi{}
structure types are T$_2$Pn$_2$ layers which are alternately stacked
with R$_2$O$_2$ or A layers in the RTPnO and AT$_2$Pn$_2$ families,
respectively.  The fact that the highest transition temperatures in
both families occur in compounds containing Fe$_2$As$_2$ layers
suggests that the T$_2$Pn$_2$ layers are the active layers while the
R$_2$O$_2$ or A layers act as a spacer that can fine tune the
electronic structure of the T$_2$Pn$_2$ layer and act as a charge
reservoir layer, but do not control the physics. Since
superconductivity has been found in
LaNiAsO\cite{Watanabe2008LaNiAsO}, one might expect that
superconductivity would also be found in the \ThCrSi{} structure
type with an active Ni$_2$As$_2$ layer.

Following this reasoning, we have synthesized single crystals of
\BaNiAs. We find a first order phase transition at $T_0$ = 130 K
(cooling) with 7 K thermal hysteresis. By analogy with \AFeAs (A =
Ba, Sr,
Ca)\cite{Rotter2008a,Wu2008Ca,RonningJPCM2008CaFe2As2,Ni2008b,Krellner2008Sr}
we identify this transition as a magnetic spin density wave (SDW)
transition concomitant with a structural transition. Here we show
that \BaNiAs~is also a bulk superconductor at $T_c$ = 0.7 K, well
below the first order phase transition $T_0$.


Single crystals of BaNi$_2$As$_2$ were grown in Pb flux in the ratio
Ba:Ni:As:Pb=1:2:2:20. The starting elements were placed in an
alumina crucible and sealed under vacuum in a quartz ampoule. The
ampoule was placed in a furnace and heated to 600 $^{\circ}$C at 100
$^{\circ}$C hr$^{-1}$, and held at that temperature for 4 hours.
This sequence was repeated at 900$^{\circ}$C and at a maximum
temperature of 1075 $^{\circ}$C, with hold times of 4 hr, each.  The
sample was then cooled slowly ($\sim7 ^{\circ}$C hr$^{-1}$) to 650
$^{\circ}$C, at which point the excess Pb flux was removed with the
aid of a centrifuge. The resulting plate-like crystals of typical
dimensions 1 x 1 x 0.1 mm$^3$ are micaceous and air sensitive and
are oriented with the $c$-axis normal to the plate. \BaNiAs{}
crystallizes in the ThCr$_2$Si$_2$ tetragonal structure (space group
no. 139). Single crystal refinement [R(I$>$2$\sigma$) = 5.37\%] at
room temperature gives lattice parameters $a$ = 4.112(4) \AA{} and
$c$ = 11.54(2) \AA{} and fully occupied atomic positions Ba
2a(0,0,0), Ni 4d(0.5,0,0.5) and As 4e(0,0,z) with z = 0.3476(3)
consistent with previous reports\cite{Pfisterer1980,Pfisterer1983}.
Powder X-ray diffraction data was consistent with the single crystal
diffraction data.


Specific heat measurements were carried out using an adiabatic
relaxation method in a commercial cryostat from 2 K to 300 K, and in
a dilution refrigerator down to 150 mK. Electrical transport
measurements were performed using a LR-700 resistance bridge with an
excitation current of 0.2 mA, on samples for which platinum leads
were spot welded. X-ray data were collected at room temperature on a
Bruker APEXII diffractometer, with charge-coupled-device detector,
and graphite monchromated MoK$_{\alpha}$ ($\lambda$ = 0.71073 {\AA})
radiation.  The data were corrected for absorption and
Lorentz-polarization effects..

\begin{figure}
\includegraphics[width=3.3in]{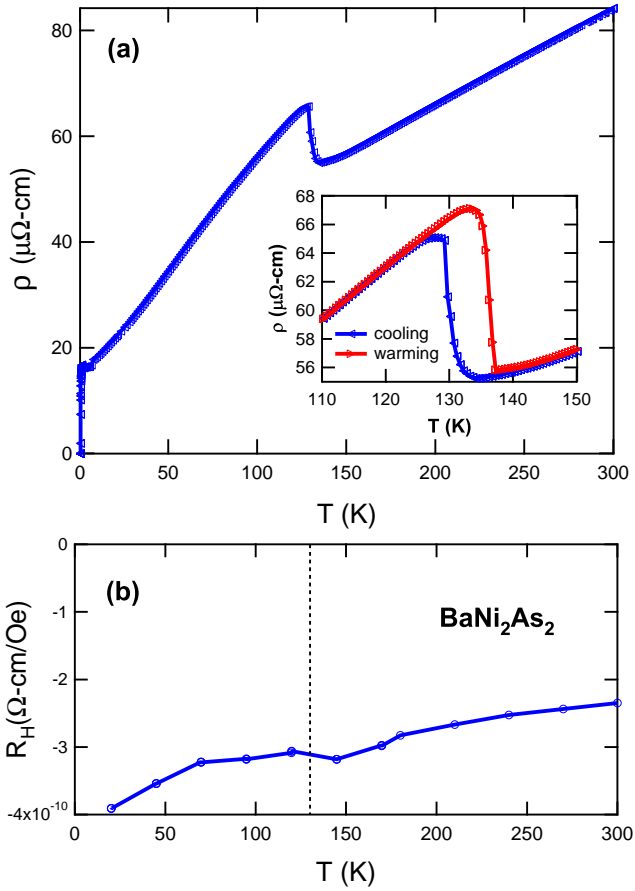}
\caption{\label{BaNiAsRes}(color online) Transport properties of
\BaNiAs. (a) In-plane resistivity (I$\parallel$ab). The inset
demonstrates the thermal hysteresis at the transition. (b) Hall
coefficient $R_H$ of \BaNiAs{}. The dashed
     line indicates the first order transition temperature.}
\end{figure}

\begin{figure}
\includegraphics[width=3.3in]{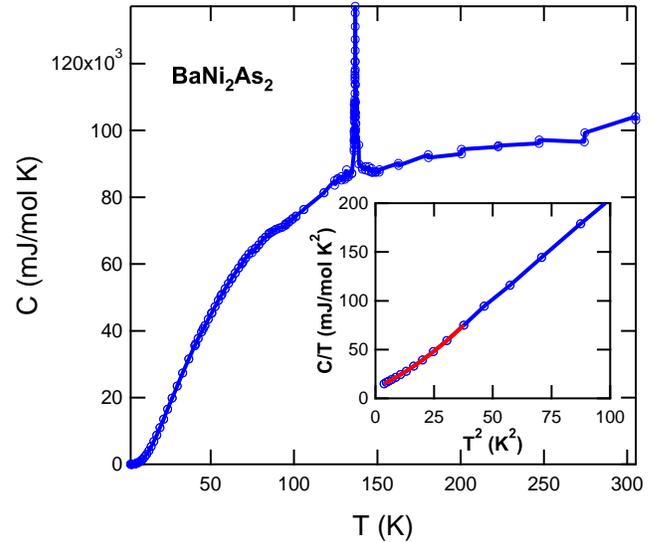}
\caption{\label{Cp}(color online) Specific heat versus temperature
is shown for BaNi$_2$As$_2$. The inset displays the low temperature
heat capacity. The solid line is a fit to $C/T = \gamma + \beta T^2
+ \alpha T^4$. }
\end{figure}

Resistivity and heat capacity shown in figures \ref{BaNiAsRes} and
\ref{Cp}, respectively, provide clear evidence for a first order
transition which occurs at 130 K upon cooling, and at 137 K upon
warming, consistent with earlier magnetic susceptibility results on
polycrystaline samples\cite{Pfisterer1983}. The resistivity anomaly
is very similar to that observed in
\CaFeAs{}\cite{Wu2008Ca,RonningJPCM2008CaFe2As2,Ni2008b} with a RRR
(= $\rho$(300 K)/$\rho$(4 K)) of 5, while the absolute magnitude of
the resistivity is more than an order of magnitude less than in the
\AFeAs{} compounds. The thermal hysteresis of 7 K is clearly
observed in the resistivity data shown as an inset to figure
\ref{BaNiAsRes}. The Hall coefficient is negative over the entire
temperature range, and indicates a weak anomaly at $T_0$. The value
at 2 K is $R_H$ = -4x10$^{-10}$ $\Omega$-cm/Oe. The sharp anomaly at
137 K in the heat capacity data of figure \ref{Cp} (taken upon
warming) is also consistent with a first order phase transition.
From 2 to 6 K the heat capacity data was fit to $C = \gamma T +
\beta T^3 + \alpha T^5$. This yields a Sommerfeld coefficient
$\gamma$ = 10.8 $\pm$ 0.1 mJ/mol K$^2$. Assuming that the $T^3$ term
is due solely to acoustic phonons, the $\beta$ coefficient = 1.10
$\pm$ 0.01 mJ/mol K$^4$ gives a Debye temperature $\Theta_D$ = 206
K.

\begin{figure}[htbp]
     \centering
     \includegraphics[width=0.5\textwidth]{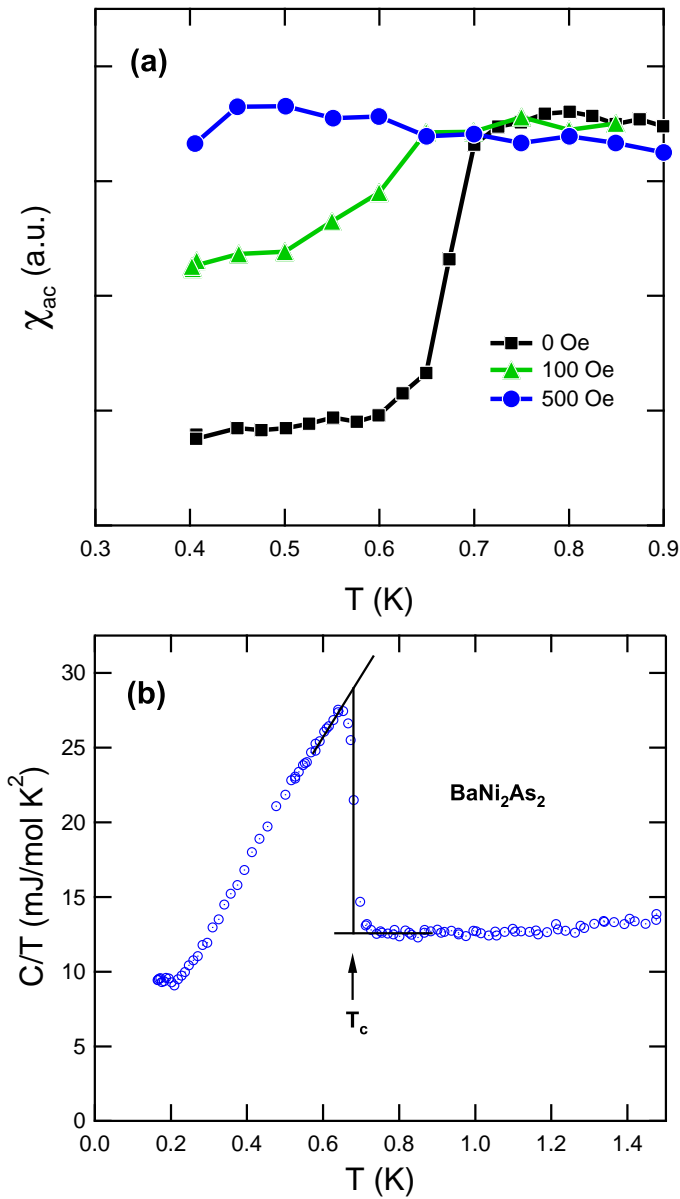}
     \caption{(color online) (a) ac magnetic susceptibility $\chi_{ac}(T)$ of \BaNiAs{} with
     applied magnetic field. (b) Low temperature specific heat of
     \BaNiAs{}.}
\label{acChi}
 \end{figure}

At low temperatures, ac-susceptibility, heat capacity, and
resistivity provide evidence for bulk superconductivity. As shown in
figure \ref{acChi}a the onset of diamagnetism starts at 0.7 K and is
estimated to be $>$ 50\% volume fraction, by comparing the signal to
that of a piece of Pb with a comparable volume. The low temperature
heat capacity data on a second sample shown in figure \ref{acChi}b
reveals a sharp anomaly at 0.68 K with a jump $\Delta{}C$ = 11.15
mJ/mol K. Taking the value of the Sommerfeld coefficient at $T_c$
($\gamma$ = 12.5 mJ/mol K$^2$\cite{lowTCpfootnote}) gives the ratio
$\Delta{}C/\gamma{}T_c$ = 1.31. The large ratio confirms the bulk
nature of superconductivity, but further work is necessary to
determine whether the heat capacity data can reveal any sign of
unconventional superconducting behavior.

\begin{figure}[htbp]
     \centering
     \includegraphics[width=0.5\textwidth]{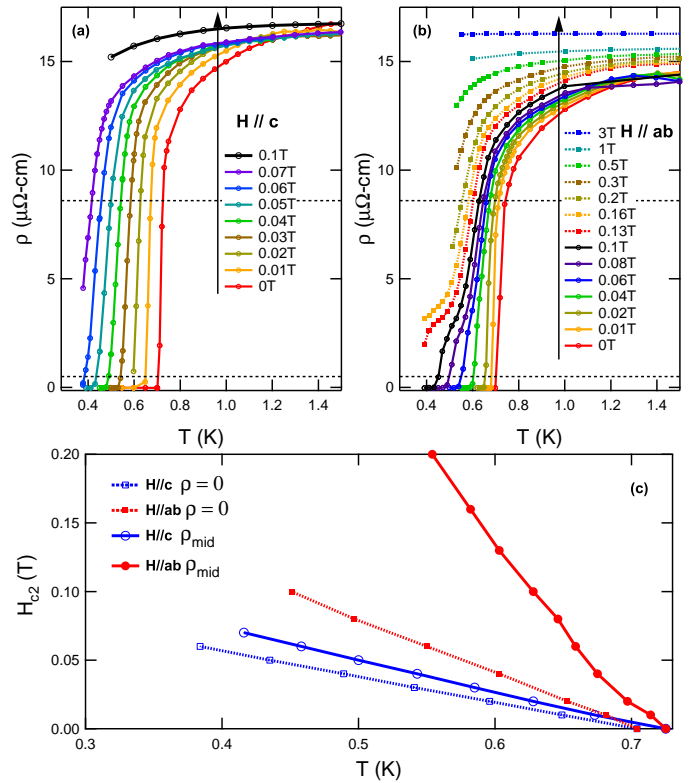}
     \caption{(color online) Evidence for superconductivity from
     the resistive transition for $H \parallel \hat{c}$ (a) and
     $H \parallel \hat{ab}$ (b). (c) Upper critical field for
     $H \parallel \hat{c}$ and
     $H \parallel \hat{ab}$ as determined by the dashed lines in
     panels (a) and (b). The current was always perpendicular to the
     magnetic field.}
\label{ResSC}
 \end{figure}

From the resistivity data at low temperatures shown in figure
\ref{ResSC} we can extract additional information. The resistivity
sample has trace amounts of Pb impurities which gives a partial
transition at 7.2 K. At roughly 1.5 K there is an additional
downturn in the resistivity data, which then goes to zero abruptly
at 0.7 K. Since the bulk transition occurs sharply at 0.7 K in zero
field, we attribute the downturn at 1.5 K to an unknown impurity
phase which is also superconducting. Upon application of a magnetic
field, we estimate the upper critical field for both the bulk
superconductor and the impurity phase. We extract the upper critical
field, $H_{c2}(T)$ for \BaNiAs{} by taking the temperature at which
$\rho$ = 0.5 $\mu\Omega$-cm (the lower dashed line in figure
\ref{ResSC}a and \ref{ResSC}b). This gives initial slopes of
$dH_{c2}^{ab}/dT$ = -0.396 T/K and $dH_{c2}^{c}/dT$ = -0.186 T/K
with an anisotropy of 2.1. From these initial slopes we estimate the
zero-temperature upper critical field $H_{c2}$(0) =
-0.7$T_cdH_{c2}/dT_c$ \cite{WHH1966} to be 0.19 T and 0.09 T for
$H\parallel{}ab$ and $H\parallel{}c$, respectively, yielding a
Ginzburg-Landau coherence length $\xi^{ab}$ = 420 {\AA} and
$\xi^{c}$ = 610 {\AA}, using the formula $\xi$ =
($\Phi/2\pi{}H_{c2}(0)$)$^{1/2}$, where $\Phi$ = 2.07 $\times$
10$^{-7}$ Oe cm$^2$ is the flux quantum. Surprisingly, for the
magnetic field in the $ab$-plane the resistive anomaly develops a
shoulder. Consequently, the upper critical field of the impurity
phase, for which we obtain a rough estimate by taking the midpoint
of the resistive transition, has even greater anisotropy than the
bulk \BaNiAs{} superconductor.

Whether superconductivity can coexist with the low temperature
orthorhombic structure and/or the spin-density wave (SDW) ground
state is unclear. While coexistence of SDW and SC order is observed
in the phase diagram of some doped compounds (e.g. ref. 2), whether
or not this is microscopic coexistence remains to be determined. Of
the stoichiometric compounds which superconduct in either the
ZrCuSiAs\cite{Watanabe2007LaNiPO,Watanabe2008LaNiAsO,Li2008NiAs} or
\ThCrSi
\cite{Park2008CaFe2As2,Torikachvili2008CaFe2As2,Mine2008BaNi2P2,
Jeitschko1987LaRu2P2,Sasmal2008} structure, to our knowledge, none
have yet been shown to coexist with a magnetic ground state.
Microscopic confirmation of a low temperature orthorhombic
possessing a spin-density wave is still needed in \BaNiAs. However,
the similarity of the first order anomaly here to those found in the
\AFeAs{} systems
\cite{Rotter2008a,Krellner2008Sr,RonningJPCM2008CaFe2As2,Ni2008b,Wu2008Ca}
where an orthorhombic SDW state has been determined
\cite{Huang2008BaNeutrons,Jesche2008SrNeutron} is suggestive that a
similar situation occurs in \BaNiAs. Thus, the clear observation of
bulk superconductivity below the first order transition in \BaNiAs,
may constitute the first example of coexistence of these three order
parameters in a system with active T$_2$Pn$_2$ layers.

The observation of bulk superconductivity at 0.7 K in \BaNiAs{}
completes a form of continuity with regards to the presence of
superconductivity in going from the ZrCuSiAs structure type to the
\ThCrSi{} structure type, independent of whether the active layers
are Fe$_2$As$_2$, Ni$_2$P$_2$, or Ni$_2$As$_2$. When the active
layers are Fe$_2$As$_2$, the stoichiometric materials possess SDW
order, and require doping or pressure to produce superconductivity.
In the cases of Ni$_2$P$_2$ and Ni$_2$As$_2$ layers, the
stoichiometric parent compounds possess superconductivity in both
structure types. The biggest difference with these comparisons is
that \BaNiAs has a first order phase transition, while LaNiAsO does
not.

In conclusion, we have synthesized single crystals of
BaNi$_2$As$_2$, which possesses both a first order transition at 130
K, which is likely a combined structural and magnetic transition,
and superconductivity at 0.7 K. It will be interesting to study the
dependence of doping, pressure, and isoelectronic substitution on
these transitions to help elucidate the origin of superconductivity
as well as the influence of competing orders.

\begin{acknowledgments} We thank H. Lee for assistance with the measurements.
Work at Los Alamos National Laboratory was
performed under the auspices of the U.S. Department of Energy.
\end{acknowledgments}


\begin{thebibliography}{25}
\expandafter\ifx\csname natexlab\endcsname\relax\def\natexlab#1{#1}\fi \expandafter\ifx\csname bibnamefont\endcsname\relax
  \def\bibnamefont#1{#1}\fi
\expandafter\ifx\csname bibfnamefont\endcsname\relax
  \def\bibfnamefont#1{#1}\fi
\expandafter\ifx\csname citenamefont\endcsname\relax
  \def\citenamefont#1{#1}\fi
\expandafter\ifx\csname url\endcsname\relax
  \def\url#1{\texttt{#1}}\fi
\expandafter\ifx\csname urlprefix\endcsname\relax\def\urlprefix{URL }\fi \providecommand{\bibinfo}[2]{#2} \providecommand{\eprint}[2][]{\url{#2}}

\bibitem[{\citenamefont{}(2008)}]{ZARen2008a}
\bibinfo{author}{\bibfnamefont{Z.-A.} \bibnamefont{Ren}},
\bibinfo{author}{\bibfnamefont{W.} \bibnamefont{Lu}},
\bibinfo{author}{\bibfnamefont{J.} \bibnamefont{Yang}},
\bibinfo{author}{\bibfnamefont{W.} \bibnamefont{Yi}},
\bibinfo{author}{\bibfnamefont{X-L.} \bibnamefont{Shen}},
\bibinfo{author}{\bibfnamefont{Z-C.} \bibnamefont{Li}},
\bibinfo{author}{\bibfnamefont{G-C.} \bibnamefont{Che}},
\bibinfo{author}{\bibfnamefont{X-L.} \bibnamefont{Dong}},
\bibinfo{author}{\bibfnamefont{L-L.} \bibnamefont{Sun}},
\bibinfo{author}{\bibfnamefont{F.} \bibnamefont{Zhou}},
\bibinfo{author}{\bibfnamefont{Z-X.} \bibnamefont{Zhou}},
  \bibinfo{journal}{Chin. Phys. Lett.} \textbf{\bibinfo{volume}{25}},
  \bibinfo{pages}{2215} (\bibinfo{year}{2008}).

\bibitem[{\citenamefont{Kamihara}(2008)}]{KamiharaJACS2008}
\bibinfo{author}{\bibfnamefont{Y.} \bibnamefont{Kamihara}},
\bibinfo{author}{\bibfnamefont{T.} \bibnamefont{Watanabe}},
\bibinfo{author}{\bibfnamefont{M.} \bibnamefont{Hirano}},
\bibinfo{author}{\bibfnamefont{H.} \bibnamefont{Hosono}},
  \bibinfo{journal}{J. Am. Chem. Soc.} \textbf{\bibinfo{volume}{130}},
  \bibinfo{pages}{3296} (\bibinfo{year}{2008}).

\bibitem[{\citenamefont{}(2008)}]{XHChenNature2008}
\bibinfo{author}{\bibfnamefont{X.H.} \bibnamefont{Chen}},
\bibinfo{author}{\bibfnamefont{T.} \bibnamefont{Wu}},
\bibinfo{author}{\bibfnamefont{G.} \bibnamefont{Wu}},
\bibinfo{author}{\bibfnamefont{R.H.} \bibnamefont{Liu}},
\bibinfo{author}{\bibfnamefont{H.} \bibnamefont{Chen}},
\bibinfo{author}{\bibfnamefont{D.F.} \bibnamefont{Fang}},
  \bibinfo{journal}{Nature} \textbf{\bibinfo{volume}{453}},
  \bibinfo{pages}{761} (\bibinfo{year}{2008}).


%
\bibitem[{\citenamefont{}(2008)}]{GFChen2008a}
\bibinfo{author}{\bibfnamefont{G.F.} \bibnamefont{Chen}},
\bibinfo{author}{\bibfnamefont{Z.} \bibnamefont{Li}},
\bibinfo{author}{\bibfnamefont{D.} \bibnamefont{Wu}},
\bibinfo{author}{\bibfnamefont{G.} \bibnamefont{Li}},
\bibinfo{author}{\bibfnamefont{W.Z.} \bibnamefont{Hu}},
\bibinfo{author}{\bibfnamefont{J.} \bibnamefont{Dong}},
\bibinfo{author}{\bibfnamefont{P.} \bibnamefont{Zheng}},
\bibinfo{author}{\bibfnamefont{J.L.} \bibnamefont{Luo}},
\bibinfo{author}{\bibfnamefont{N.L.} \bibnamefont{Wang}},
  \bibinfo{journal}{Phys. Rev. Lett.} \textbf{\bibinfo{volume}{100}},
  \bibinfo{pages}{247002} (\bibinfo{year}{2008}).

\bibitem[{\citenamefont{}(2008)}]{ZARen2008b}
\bibinfo{author}{\bibfnamefont{Z.-A.} \bibnamefont{Ren}},
\bibinfo{author}{\bibfnamefont{J.} \bibnamefont{Yang}},
\bibinfo{author}{\bibfnamefont{W.} \bibnamefont{Lu}},
\bibinfo{author}{\bibfnamefont{W.} \bibnamefont{Yi}},
\bibinfo{author}{\bibfnamefont{X-L.} \bibnamefont{Shen}},
\bibinfo{author}{\bibfnamefont{Z-C.} \bibnamefont{Li}},
\bibinfo{author}{\bibfnamefont{G-C.} \bibnamefont{Che}},
\bibinfo{author}{\bibfnamefont{X-L.} \bibnamefont{Dong}},
\bibinfo{author}{\bibfnamefont{L-L.} \bibnamefont{Sun}},
\bibinfo{author}{\bibfnamefont{F.} \bibnamefont{Zhou}},
\bibinfo{author}{\bibfnamefont{Z-X.} \bibnamefont{Zhou}},
  \bibinfo{journal}{Europhys. Lett.} \textbf{\bibinfo{volume}{82}},
  \bibinfo{pages}{57002} (\bibinfo{year}{2008}).

\bibitem[{\citenamefont{}(2008)}]{ZARen2008c}
\bibinfo{author}{\bibfnamefont{Z.-A.} \bibnamefont{Ren}},
\bibinfo{author}{\bibfnamefont{J.} \bibnamefont{Yang}},
\bibinfo{author}{\bibfnamefont{W.} \bibnamefont{Lu}},
\bibinfo{author}{\bibfnamefont{W.} \bibnamefont{Yi}},
\bibinfo{author}{\bibfnamefont{G-C.} \bibnamefont{Che}},
\bibinfo{author}{\bibfnamefont{X-L.} \bibnamefont{Dong}},
\bibinfo{author}{\bibfnamefont{L-L.} \bibnamefont{Sun}},
\bibinfo{author}{\bibfnamefont{Z-X.} \bibnamefont{Zhou}},
  \bibinfo{journal}{Mater. Res. Innov.} \textbf{\bibinfo{volume}{12}},
  \bibinfo{pages}{1} (\bibinfo{year}{2008}).

\bibitem[{\citenamefont{}(2008)}]{PCheng2008}
\bibinfo{author}{\bibfnamefont{P.} \bibnamefont{Cheng}},
\bibinfo{author}{\bibfnamefont{L.} \bibnamefont{Fang}},
\bibinfo{author}{\bibfnamefont{H.} \bibnamefont{Yang}},
\bibinfo{author}{\bibfnamefont{X.} \bibnamefont{Zhu}},
\bibinfo{author}{\bibfnamefont{G.} \bibnamefont{Mu}},
\bibinfo{author}{\bibfnamefont{H.} \bibnamefont{Luo}},
\bibinfo{author}{\bibfnamefont{Z.} \bibnamefont{Wang}},
\bibinfo{author}{\bibfnamefont{H.-H.} \bibnamefont{Wen}},
  \bibinfo{journal}{Science in China G} \textbf{\bibinfo{volume}{51}},
  \bibinfo{pages}{719} (\bibinfo{year}{2008}).

\bibitem[{\citenamefont{}(2007)}]{Watanabe2007LaNiPO}
\bibinfo{author}{\bibfnamefont{T.} \bibnamefont{Watanabe}},
\bibinfo{author}{\bibfnamefont{H.} \bibnamefont{Yanagi}},
\bibinfo{author}{\bibfnamefont{T.} \bibnamefont{Kamiya}},
\bibinfo{author}{\bibfnamefont{Y.} \bibnamefont{Kamihara}},
\bibinfo{author}{\bibfnamefont{H.} \bibnamefont{Hiramatsu}},
\bibinfo{author}{\bibfnamefont{M.} \bibnamefont{Hirano}},
\bibinfo{author}{\bibfnamefont{H.} \bibnamefont{Hosono}},
  \bibinfo{journal}{Inorg. Chem.} \textbf{\bibinfo{volume}{46}},
  \bibinfo{pages}{7719} (\bibinfo{year}{2007}).

\bibitem[{\citenamefont{}(2008)}]{Watanabe2008LaNiAsO}
\bibinfo{author}{\bibfnamefont{T.} \bibnamefont{Watanabe}},
\bibinfo{author}{\bibfnamefont{H.} \bibnamefont{Yanagi}},
\bibinfo{author}{\bibfnamefont{Y.} \bibnamefont{Kamihara}},
\bibinfo{author}{\bibfnamefont{T.} \bibnamefont{Kamiya}},
\bibinfo{author}{\bibfnamefont{M.} \bibnamefont{Hirano}},
\bibinfo{author}{\bibfnamefont{H.} \bibnamefont{Hosono}},
  \bibinfo{journal}{J. Solid State Chem.} \textbf{\bibinfo{volume}{}},
  \bibinfo{pages}{doi:10.1016/j.jssc.2008.04.033} (\bibinfo{year}{2008}).

\bibitem[{\citenamefont{}(2008)}]{Fang2008NiAs}
\bibinfo{author}{\bibfnamefont{L.} \bibnamefont{Fang}},
\bibinfo{author}{\bibfnamefont{H.} \bibnamefont{Yang}},
\bibinfo{author}{\bibfnamefont{P.} \bibnamefont{Cheng}},
\bibinfo{author}{\bibfnamefont{X.} \bibnamefont{Zhu}},
\bibinfo{author}{\bibfnamefont{G.} \bibnamefont{Mu}},
\bibinfo{author}{\bibfnamefont{H.-H.} \bibnamefont{Wen}},
  \bibinfo{journal}{arXiv:0803.3978} \textbf{\bibinfo{volume}{}},
  \bibinfo{pages}{} (\bibinfo{year}{2008}).

\bibitem[{\citenamefont{}(2008)}]{Li2008NiAs}
\bibinfo{author}{\bibfnamefont{Z.} \bibnamefont{Li}},
\bibinfo{author}{\bibfnamefont{G.F.} \bibnamefont{Chen}},
\bibinfo{author}{\bibfnamefont{J.} \bibnamefont{Dong}},
\bibinfo{author}{\bibfnamefont{G.} \bibnamefont{Li}},
\bibinfo{author}{\bibfnamefont{W.Z.} \bibnamefont{Hu}},
\bibinfo{author}{\bibfnamefont{D.} \bibnamefont{Wu}},
\bibinfo{author}{\bibfnamefont{S.K.} \bibnamefont{Su}},
\bibinfo{author}{\bibfnamefont{P.} \bibnamefont{Zheng}},
\bibinfo{author}{\bibfnamefont{T.} \bibnamefont{Xiang}},
\bibinfo{author}{\bibfnamefont{N.L.} \bibnamefont{Wang}},
\bibinfo{author}{\bibfnamefont{J.L.} \bibnamefont{Luo}},
  \bibinfo{journal}{arXiv:0803.2572} \textbf{\bibinfo{volume}{}},
  \bibinfo{pages}{} (\bibinfo{year}{2008}).

\bibitem[{\citenamefont{}(2006)}]{Kamihara2006LaFePO}
\bibinfo{author}{\bibfnamefont{Y.} \bibnamefont{Kamihara}},
\bibinfo{author}{\bibfnamefont{H.} \bibnamefont{Hiramatsu}},
\bibinfo{author}{\bibfnamefont{M.} \bibnamefont{Hirano}},
\bibinfo{author}{\bibfnamefont{R.} \bibnamefont{Kawamura}},
\bibinfo{author}{\bibfnamefont{H.} \bibnamefont{Yanagi}},
\bibinfo{author}{\bibfnamefont{T.} \bibnamefont{Kamiya}},
\bibinfo{author}{\bibfnamefont{H.} \bibnamefont{Hosono}},
  \bibinfo{journal}{J. Am. Chem. Soc.} \textbf{\bibinfo{volume}{128}},
  \bibinfo{pages}{10012} (\bibinfo{year}{2006}).

\bibitem[{\citenamefont{Rotter}(2008)}]{Rotter2008b}
\bibinfo{author}{\bibfnamefont{M.} \bibnamefont{Rotter}},
\bibinfo{author}{\bibfnamefont{M.} \bibnamefont{Tegel}},
\bibinfo{author}{\bibfnamefont{D.} \bibnamefont{Johrendt}},
  \bibinfo{journal}{arXiv:0805.4630} \textbf{\bibinfo{volume}{}},
  \bibinfo{pages}{} (\bibinfo{year}{2008}).

\bibitem[{\citenamefont{Chen}(2008)}]{GFChen2008b}
\bibinfo{author}{\bibfnamefont{G.F.} \bibnamefont{Chen}},
\bibinfo{author}{\bibfnamefont{Z.} \bibnamefont{Li}},
\bibinfo{author}{\bibfnamefont{G.} \bibnamefont{Li}},
\bibinfo{author}{\bibfnamefont{W.Z.} \bibnamefont{Hu}},
\bibinfo{author}{\bibfnamefont{J.} \bibnamefont{Dong}},
\bibinfo{author}{\bibfnamefont{X.D.} \bibnamefont{Zhang}},
\bibinfo{author}{\bibfnamefont{P.} \bibnamefont{Zheng}},
\bibinfo{author}{\bibfnamefont{N.L.} \bibnamefont{Wang}},
\bibinfo{author}{\bibfnamefont{J.L.} \bibnamefont{Luo}},
  \bibinfo{journal}{arXiv:0806.1209} \textbf{\bibinfo{volume}{}},
  \bibinfo{pages}{} (\bibinfo{year}{2008}).

\bibitem[{\citenamefont{Sasmal}(2008)}]{Sasmal2008}
\bibinfo{author}{\bibfnamefont{K.} \bibnamefont{Sasmal}},
\bibinfo{author}{\bibfnamefont{B.} \bibnamefont{Lv}},
\bibinfo{author}{\bibfnamefont{B.} \bibnamefont{Lorenz}},
\bibinfo{author}{\bibfnamefont{A.} \bibnamefont{Guloy}},
\bibinfo{author}{\bibfnamefont{F.} \bibnamefont{Chen}},
\bibinfo{author}{\bibfnamefont{Y.} \bibnamefont{Xue}},
\bibinfo{author}{\bibfnamefont{C.W.} \bibnamefont{Chu}},
  \bibinfo{journal}{arXiv:0806.1301} \textbf{\bibinfo{volume}{}},
  \bibinfo{pages}{} (\bibinfo{year}{2008}).

\bibitem[{\citenamefont{}(2008)}]{Wu2008Ca}
\bibinfo{author}{\bibfnamefont{G.} \bibnamefont{Wu}},
\bibinfo{author}{\bibfnamefont{H.} \bibnamefont{Chen}},
\bibinfo{author}{\bibfnamefont{T.} \bibnamefont{Wu}},
\bibinfo{author}{\bibfnamefont{Y.L.} \bibnamefont{Xie}},
\bibinfo{author}{\bibfnamefont{Y.J.} \bibnamefont{Yan}},
\bibinfo{author}{\bibfnamefont{R.H.} \bibnamefont{Liu}},
\bibinfo{author}{\bibfnamefont{X.F.} \bibnamefont{Wang}},
\bibinfo{author}{\bibfnamefont{J.J.} \bibnamefont{Ying}},
\bibinfo{author}{\bibfnamefont{X.H.} \bibnamefont{Chen}},
  \bibinfo{journal}{arXiv:0806.4279} \textbf{\bibinfo{volume}{}},
  \bibinfo{pages}{} (\bibinfo{year}{2008}).

\bibitem[{\citenamefont{Jeevan}(2008)}]{Jeevan2008EuFe2As2SC}
\bibinfo{author}{\bibfnamefont{H.S.} \bibnamefont{Jeevan}},
\bibinfo{author}{\bibfnamefont{Z.} \bibnamefont{Hossain}},
\bibinfo{author}{\bibfnamefont{C.} \bibnamefont{Geibel}},
\bibinfo{author}{\bibfnamefont{P.} \bibnamefont{Gegenwart}},
  \bibinfo{journal}{arXiv:0807.2530} \textbf{\bibinfo{volume}{}},
  \bibinfo{pages}{} (\bibinfo{year}{2008}).

\bibitem[{\citenamefont{}(2008)}]{Park2008CaFe2As2}
\bibinfo{author}{\bibfnamefont{T.} \bibnamefont{Park}},
\bibinfo{author}{\bibfnamefont{E.} \bibnamefont{Park}},
\bibinfo{author}{\bibfnamefont{H.} \bibnamefont{Lee}},
\bibinfo{author}{\bibfnamefont{T.} \bibnamefont{Klimczuk}},
\bibinfo{author}{\bibfnamefont{E.D.} \bibnamefont{Bauer}},
\bibinfo{author}{\bibfnamefont{F.} \bibnamefont{Ronning}},
\bibinfo{author}{\bibfnamefont{J.D.} \bibnamefont{Thompson}},
  \bibinfo{journal}{J. Phys. Cond. Matter} \textbf{\bibinfo{volume}{20}},
  \bibinfo{pages}{322204} (\bibinfo{year}{2008}).

\bibitem[{\citenamefont{}(2008)}]{Torikachvili2008CaFe2As2}
\bibinfo{author}{\bibfnamefont{M.S.} \bibnamefont{Torikachvili}},
\bibinfo{author}{\bibfnamefont{S.L.} \bibnamefont{Budko}},
\bibinfo{author}{\bibfnamefont{N.} \bibnamefont{Ni}},
\bibinfo{author}{\bibfnamefont{P.C.} \bibnamefont{Canfield}},
  \bibinfo{journal}{} \textbf{\bibinfo{volume}{}},
  \bibinfo{pages}{arXiv:0807.0616} (\bibinfo{year}{2008}).

\bibitem[{\citenamefont{}(2008)}]{Alireza2008BaFe2As2}
\bibinfo{author}{\bibfnamefont{P.L.} \bibnamefont{Alireza}},
\bibinfo{author}{\bibfnamefont{J.} \bibnamefont{Gillet}},
\bibinfo{author}{\bibfnamefont{Y.T.} \bibnamefont{Chris Ko}},
\bibinfo{author}{\bibfnamefont{S.E.} \bibnamefont{Sebastian}},
\bibinfo{author}{\bibfnamefont{G.G.} \bibnamefont{Lonzarich}},
  \bibinfo{journal}{} \textbf{\bibinfo{volume}{}},
  \bibinfo{pages}{arXiv:0807.1896} (\bibinfo{year}{2008}).

\bibitem[{\citenamefont{}(2008)}]{Mine2008BaNi2P2}
\bibinfo{author}{\bibfnamefont{T.} \bibnamefont{Mine}},
\bibinfo{author}{\bibfnamefont{H.} \bibnamefont{Yanagi}},
\bibinfo{author}{\bibfnamefont{T.} \bibnamefont{Kamiya}},
\bibinfo{author}{\bibfnamefont{Y.} \bibnamefont{Kamihara}},
\bibinfo{author}{\bibfnamefont{M.} \bibnamefont{Hirano}},
\bibinfo{author}{\bibfnamefont{H.} \bibnamefont{Hosono}},
  \bibinfo{journal}{Solid State Communications} \textbf{\bibinfo{volume}{147}},
  \bibinfo{pages}{111} (\bibinfo{year}{2008}).

\bibitem[{\citenamefont{}(2008)}]{Jeitschko1987LaRu2P2}
\bibinfo{author}{\bibfnamefont{W.} \bibnamefont{Jeitschko}},
\bibinfo{author}{\bibfnamefont{R.} \bibnamefont{Glaum}},
\bibinfo{author}{\bibfnamefont{L.} \bibnamefont{Boonk}},
  \bibinfo{journal}{J. Solid State Chem.} \textbf{\bibinfo{volume}{69}},
  \bibinfo{pages}{93} (\bibinfo{year}{1987}).

\bibitem[{\citenamefont{Rotter}(2008)}]{Rotter2008a}
\bibinfo{author}{\bibfnamefont{M.} \bibnamefont{Rotter}},
\bibinfo{author}{\bibfnamefont{M.} \bibnamefont{Tegel}},
\bibinfo{author}{\bibfnamefont{D.} \bibnamefont{Johrendt}},
\bibinfo{author}{\bibfnamefont{I.} \bibnamefont{Schellenberg}},
\bibinfo{author}{\bibfnamefont{W.} \bibnamefont{Hermes}},
\bibinfo{author}{\bibfnamefont{R.} \bibnamefont{Poettgen}},
  \bibinfo{journal}{Phys. Rev. B} \textbf{\bibinfo{volume}{78}},
  \bibinfo{pages}{020503} (\bibinfo{year}{2008}).

\bibitem[{\citenamefont{}(2008)}]{Krellner2008Sr}
\bibinfo{author}{\bibfnamefont{C.} \bibnamefont{Krellner}},
\bibinfo{author}{\bibfnamefont{N.} \bibnamefont{Caroca-Canales}},
\bibinfo{author}{\bibfnamefont{A.} \bibnamefont{Jesche}},
\bibinfo{author}{\bibfnamefont{H.} \bibnamefont{Rosner}},
\bibinfo{author}{\bibfnamefont{A.} \bibnamefont{Ormeci}},
\bibinfo{author}{\bibfnamefont{C.} \bibnamefont{Geibel}},
  \bibinfo{journal}{arXiv:0806.1043} \textbf{\bibinfo{volume}{}},
  \bibinfo{pages}{} (\bibinfo{year}{2008}).

\bibitem[{\citenamefont{}(2008)}]{RonningJPCM2008CaFe2As2}
\bibinfo{author}{\bibfnamefont{F.} \bibnamefont{Ronning}},
\bibinfo{author}{\bibfnamefont{T.} \bibnamefont{Klimczuk}},
\bibinfo{author}{\bibfnamefont{E.D.} \bibnamefont{Bauer}},
\bibinfo{author}{\bibfnamefont{H.} \bibnamefont{Volz}},
\bibinfo{author}{\bibfnamefont{J.D.} \bibnamefont{Thompson}},
  \bibinfo{journal}{J. Phys.: Condens. Matter} \textbf{\bibinfo{volume}{20}},
  \bibinfo{pages}{322201} (\bibinfo{year}{2008}).

\bibitem[{\citenamefont{}(2008)}]{Ni2008b}
\bibinfo{author}{\bibfnamefont{N.} \bibnamefont{Ni}},
\bibinfo{author}{\bibfnamefont{S.} \bibnamefont{Nandi}},
\bibinfo{author}{\bibfnamefont{A.} \bibnamefont{Kreyssig}},
\bibinfo{author}{\bibfnamefont{A.I.} \bibnamefont{Goldman}},
\bibinfo{author}{\bibfnamefont{E.D.} \bibnamefont{Mun}},
\bibinfo{author}{\bibfnamefont{S.L.} \bibnamefont{Budko}},
\bibinfo{author}{\bibfnamefont{P.C.} \bibnamefont{Canfield}},
  \bibinfo{journal}{arXiv:0806.4328} \textbf{\bibinfo{volume}{}},
  \bibinfo{pages}{} (\bibinfo{year}{2008}).

\bibitem[{\citenamefont{}(1980)}]{Pfisterer1980}
\bibinfo{author}{\bibfnamefont{M.} \bibnamefont{Pfisterer}},
\bibinfo{author}{\bibfnamefont{G.} \bibnamefont{Nagorsen}},
  \bibinfo{journal}{Z. Naturforsch. B: Chem. Sci.} \textbf{\bibinfo{volume}{35}},
  \bibinfo{pages}{703} (\bibinfo{year}{1980}).

\bibitem[{\citenamefont{}(1983)}]{Pfisterer1983}
\bibinfo{author}{\bibfnamefont{M.} \bibnamefont{Pfisterer}},
\bibinfo{author}{\bibfnamefont{G.} \bibnamefont{Nagorsen}},
  \bibinfo{journal}{Z. Naturforsch. B: Chem. Sci.} \textbf{\bibinfo{volume}{38}},
  \bibinfo{pages}{811} (\bibinfo{year}{1983}).

\bibitem[{\citenamefont{}()}]{lowTCpfootnote}
  \bibinfo{journal}{We attribute the difference in $\gamma$ obtained from the data
  at $T_c$ and the extrapolation from higher temperature to the difficulty
  in subtracting the addenda at low temperatures.}

\bibitem[{\citenamefont{}(1966)}]{WHH1966}
\bibinfo{author}{\bibfnamefont{N.R.} \bibnamefont{Werthamer}},
\bibinfo{author}{\bibfnamefont{E.} \bibnamefont{Helfand}},
\bibinfo{author}{\bibfnamefont{P.C.} \bibnamefont{Hohenberg}},
  \bibinfo{journal}{Phys. Rev.} \textbf{\bibinfo{volume}{147}},
  \bibinfo{pages}{295} (\bibinfo{year}{1966}).

\bibitem[{\citenamefont{}(2008)}]{Huang2008BaNeutrons}
\bibinfo{author}{\bibfnamefont{Q.} \bibnamefont{Huang}},
\bibinfo{author}{\bibfnamefont{Y.} \bibnamefont{Qiu}},
\bibinfo{author}{\bibfnamefont{W.} \bibnamefont{Bao}},
\bibinfo{author}{\bibfnamefont{J.W.} \bibnamefont{Lynn}},
\bibinfo{author}{\bibfnamefont{M.A.} \bibnamefont{Green}},
\bibinfo{author}{\bibfnamefont{Y.} \bibnamefont{Chen}},
\bibinfo{author}{\bibfnamefont{T.} \bibnamefont{Wu}},
\bibinfo{author}{\bibfnamefont{G.} \bibnamefont{Wu}},
\bibinfo{author}{\bibfnamefont{X.H.} \bibnamefont{Chen}},
  \bibinfo{journal}{arXiv:0806.2776} \textbf{\bibinfo{volume}{}},
  \bibinfo{pages}{} (\bibinfo{year}{2008}).

\bibitem[{\citenamefont{}(2008)}]{Jesche2008SrNeutron}
\bibinfo{author}{\bibfnamefont{A.} \bibnamefont{Jesche}},
\bibinfo{author}{\bibfnamefont{N.} \bibnamefont{Caroca-Canales}},
\bibinfo{author}{\bibfnamefont{H.} \bibnamefont{Rosner}},
\bibinfo{author}{\bibfnamefont{H.} \bibnamefont{Borrmann}},
\bibinfo{author}{\bibfnamefont{A.} \bibnamefont{Ormeci}},
\bibinfo{author}{\bibfnamefont{D.} \bibnamefont{Kasinathan}},
\bibinfo{author}{\bibfnamefont{K.} \bibnamefont{Kaneko}},
\bibinfo{author}{\bibfnamefont{H.H.} \bibnamefont{Klauss}},
\bibinfo{author}{\bibfnamefont{H.} \bibnamefont{Luetkens}},
\bibinfo{author}{\bibfnamefont{R.} \bibnamefont{Khasanov}},
\bibinfo{author}{\bibfnamefont{A.} \bibnamefont{Amato}},
\bibinfo{author}{\bibfnamefont{A.} \bibnamefont{Hoser}},
\bibinfo{author}{\bibfnamefont{C.} \bibnamefont{Krellner}},
\bibinfo{author}{\bibfnamefont{C.} \bibnamefont{Geibel}},
  \bibinfo{journal}{arXiv:0807.0632} \textbf{\bibinfo{volume}{}},
  \bibinfo{pages}{} (\bibinfo{year}{2008}).





\end{thebibliography}
\end{document}